\documentclass[pre,nofootinbib,superscriptaddress,twocolumn]{revtex4}

\usepackage{graphicx}
\usepackage{amssymb,amsfonts,amsmath}
\usepackage{color}
\usepackage{ulem}

\def\R{{\mathbb R}}
\def\C{{\mathbb C}}

\def\x{{\mathbf x}}

\def\erfc{\mathrm{erfc}}
\def\erfcx{\mathrm{erfcx}}
\def\Ei{\mathrm{Ei}}

\begin{document}

\title{\textsf{Strong defocusing of molecular reaction times results from an interplay of geometry and reaction
control}}

\author{Denis~S.~Grebenkov}
\email{denis.grebenkov@polytechnique.edu}
\affiliation{Laboratoire de Physique de la Mati\`{e}re Condens\'{e}e (UMR 7643),\\ 
CNRS -- Ecole Polytechnique, University Paris-Saclay, 91128 Palaiseau, France}
\author{Ralf Metzler}
\affiliation{Institute of Physics \& Astronomy, University of Potsdam, 14476
Potsdam-Golm, Germany}
\author{Gleb Oshanin}
\affiliation{Sorbonne Universit\'{e}, CNRS, Laboratoire de Physique Th\'eorique de
la Mati\`ere Condens\'ee (UMR CNRS 7600), 4 place Jussieu, F-752525 Paris Cedex 05,
France}

\date{\today}

\begin{abstract}
Text-book concepts of diffusion- versus kinetic-control are
well-defined for reaction-kinetics involving macroscopic
concentrations of diffusive reactants that are adequately described by
rate-constants -- the inverse of the mean-first-passage-time to the
reaction-event.  In contradistinction, an open important question is
whether the mean-first-passage-time alone is a sufficient measure for
biochemical reactions that involve nanomolar reactant concentrations.
Here, using a simple yet generic, exactly-solvable model we study the
conspiratory effect of diffusion and chemical reaction-limitations on
the full reaction-time distribution.  We show that it has a complex
structure with four distinct regimes delimited by three characteristic
time scales spanning a window of several decades.  Consequently, the
reaction-times are defocused: no unique time-scale characterises the
reaction-process, diffusion- and kinetic-control can no longer be
disentangled, and it is imperative to know the full reaction-time
distribution.  We introduce the concepts of geometry- and
reaction-control, and also quantify each regime by calculating the
corresponding reaction depth.
\end{abstract}


\maketitle

\section*{INTRODUCTION}

Reactions between chemically active molecules in condensed matter
systems are typically controlled by two factors: the diffusive search
of the species for each other
\cite{Calef83,Weiss86,Shoup82,Lindenberg} and the intrinsic reactivity
$\kappa$ associated with the probability that a reaction indeed occurs
when the particles collide with each other \cite{Hanggi90}.  For
chemical reactions involving sufficiently high concentrations of
particles, which are initially uniformly distributed in the container
or reactor such that encounters between reactive species occur more or
less uniformly in time, theories based on mean effective reaction
rates provide an adequate description of the reaction kinetics
\cite{Calef83,Weiss86,Shoup82}---apart from some singular and
well-known reaction schemes which exhibit anomalous,
fluctuation-induced kinetics under special physical conditions (see,
for instance,
\cite{Lindenberg,Krapivsky,Oshanin89a,Oshanin89b,Yuste08,Grebenkov10a,Grebenkov10b}).
Since the seminal works by Smoluchowski \cite{Smoluchowski1917} and
Collins and Kimball \cite{Collins49} a vast number of theoretical
advances have scrutinised a combined effect of both rate-controlling
factors on the mean effective rates providing a comprehensive
understanding of this effect
\cite{Calef83,Weiss86,Shoup82,Lindenberg,Sapoval94,Grebenkov06,Holcman13,Gudowska17}.
In particular, the mean reaction time is the sum of two time scales
corresponding to the inverse diffusion coefficient and the inverse
intrinsic reactivity (see equation \eqref{eq:MFPT}), such that the
influence of diffusion control and (chemical) rate control are
separable \cite{Collins49}.

For many biochemical reactions, however, the reactive species do not
exist in sufficiently abundant amounts to give rise to smooth
concentration levels.  In contrast, only small numbers of
biomolecules, released at certain prescribed positions, are often
involved in the reaction process.  Indeed, in systems such as the well
studied Lac and phage lambda repressor proteins only few to few tens
of molecules are typically present in a living biological cell,
corresponding to nanomolar concentrations.  The starting positions of
biomolecules can either be rather close to the target or relatively
far away.  Particularly in the context of the rapid search hypothesis
of gene expression it was shown that the geometric distance between
two genes, communicating with each other via signalling proteins -- is
typically kept short by design in biological cells \cite{Kolesov07},
guaranteeing higher-than-average concentrations of proteins around the
target in conjunction with fast and reliable signalling
\cite{Pulkkinen13}.  Quite generically, many intracellular processes of
signalling, regulation, infection, immune reactions, metabolism, or
transmitter release in neurons are triggered by the arrival of one or
few biomolecules to a small spatially localised region
\cite{Alberts,Snustad}.  In such cases it becomes inappropriate to
rely on mean rates, and one needs to know the whole distribution of
random reaction times, also called the first passage times to a
reaction event. Lacking a large number of molecules, reaction times
become strongly defocused such that the mean reaction time is no
longer representative and the most probable reaction time becomes
relevant.  We note that even for perfect reactions that occur
immediately upon the first encounter between two particles and have
thus infinitely large intrinsic reactivity, the mean and the most
probable first passage times can differ by orders of magnitude
\cite{Godec16a,Godec16b} and two first passage events in the same
system may be dramatically disparate \cite{Mejia11,Mattos12,Mattos14}.

For such effectively few-body reactions, most of the available
theoretical effort has been concentrated on the analysis of perfect
reactions and hence, on the impact of diffusion control only
\cite{Redner,Benichou10,Benichou14}.  In particular, in
\cite{Benichou10} it was argued that for perfect reactions the
reaction time density (RTD) can be accurately modelled as
\begin{equation}
\label{benichou}
H(t)\approx q\delta(t)+(1-q)\exp(-t/t_{\mathrm{mean}})/t_{\mathrm{mean}},
\end{equation}
where $t_{\mathrm{mean}}$ is the MFPT and $q$ is the contribution of
trajectories which arrive to the target site immediately.  Conversely,
fluctuations of the cycle completion time for enzymatic reactions, in
absence of any diffusion stage, have been quantified through the
coefficient of variation, $\gamma$, of the corresponding distribution
function of these times \cite{Moffitt10}.  Few other works
\cite{Shoup81,Hippel89,Zon05,Grebenkov17a,Grebenkov17b} analysed the
combined effect of both rate-controlling factors but solely for the
mean reaction time.  These works have shown that the effect of the
intrinsic reactivity is certainly significant and even most likely is
the dominant factor.  The question of the combined influence of both
factors on the full distribution of reaction times has been only
addressed most recently \cite{Grebenkov2018}, with the focus on the
target search kinetics in cylindrical geometries.  However, the
results of \cite{Grebenkov2018} rely on the so-called self-consistent
approximation \cite{Shoup81} and moreover, have a somewhat cumbersome
and thus less practical form.  Hence, it is highly desirable to
consider particular yet generic examples for which the RTD can be
calculated exactly and the results can be presented in a lucid,
compact and easy to use form revealing numerous insightful features
well beyond the simple approximation in equation
\eqref{benichou}.  This is clearly an appealing problem of utmost
significance for a conceptual understanding of the kinetics of
biochemical reactions.

We here focus on the conceptually and practically relevant question of
the influence of the intrinsic chemical reactivity and the initial
position of the reacting particles onto the form of the full
distribution of reaction times.  
%
We demonstrate that when the reactivity is finite and no longer
guarantees immediate reaction on mutual encounter, the defocusing of
reaction times is strongly enhanced. Remarkably, an extended plateau
of the reaction time distribution emerges due to this reaction
control, such that the reaction times turn out to be equally probable
over several orders of magnitude.
A direct consequence of the defocusing is that the contributions of
diffusion and rate effects are no longer separable---to distinguish
from the classical concepts of diffusion and kinetic control, we will
talk about geometry (initial distance) control and reaction (intrinsic
reaction rate) control, keeping in mind that the latter not only
specifies the dominant rate-controlling factor for the MFPT, but
affects the shape of the full RTD.  An exact solution for the RTD
provides us a unique opportunity to derive explicit formulae, for
arbitrary initial conditions and arbitrary values of the intrinsic
reaction constant $\kappa$ for several characteristic properties of
the distribution such as, e.g., its precise functional forms in
different asymptotic regimes, the corresponding crossover times
between these kinetic regimes, and also the reaction depths
corresponding to these time scales.  

\section*{RESULTS}

\subsection*{Mathematical model}

We consider a model involving a pair of reactive molecules: a
partially absorbing, immobile target site of radius $\rho$ within a
bounded domain of radius $R$ limited by an impenetrable boundary, and
a molecule, initially placed at some prescribed position and diffusing
with diffusivity $D$.  Once the diffusing particle hits the surface of
the target site it reacts with (binds to) the latter with a finite,
intrinsic reaction rate $\kappa$.  The reflecting outer boundary can
mimic an impenetrable cell membrane, the reaction container's surface,
or be an effective virtual frontier of the ``zone of influence'' of
the target molecule, separating it from other remotely located target
molecules.

\begin{figure*}
\includegraphics[width=18cm]{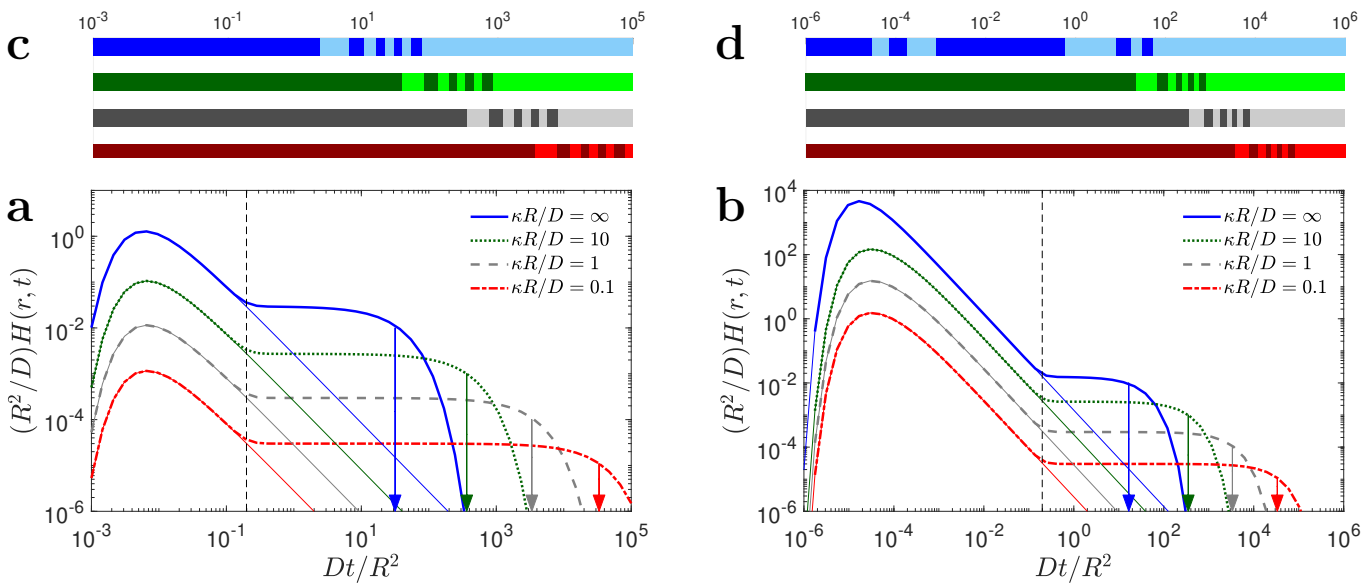}
\caption{
Reaction control.  Reaction time density $H(r,t)$ for a reaction on an
inner target of radius $\rho/R=0.01$, with starting point {\bf (a)}
$r/R=0.2$ and {\bf (b)} $r/R=0.02$ for four progressively decreasing
(from top to bottom) values of the dimensionless reactivity $\kappa' =
\kappa R/D$ indicated in the plot.  Note that $\kappa'$ includes $R$
and $D$ such that smaller values of $\kappa'$ can also be achieved at
a fixed $\kappa$ upon lowering $R$ or by increasing the values of $D$.
The coloured vertical arrows indicate the mean reaction times for
these cases.  The vertical black dashed line indicates the crossover
time $t_c=2(R-\rho)^2/(D\pi^2)$ above which the contribution of higher
order Laplacian eigenmodes become negligible.  This characteristic
time marks the end of the hump-like region (L\'evy-Smirnov region
specific to an unbounded system, see below and the Method section for
more details) and indicates the crossover to a plateau region with
equiprobable realisations of the reaction times.  This plateau region
spans a considerable window of reaction times, especially for lower
reactivity values.  Thin coloured lines show the reaction time density
$H_\infty(r,t)$ from equation \eqref{eq:Ht_Rinf} for the unbounded
case ($R\to\infty$).  Length and time units are fixed by setting $R=1$
and $R^2/D=1$.  Note the extremely broad range of relevant reaction
times (the horizontal axis) spanning over 12 orders of magnitude for
the panel {\bf (b)}.  Coloured bar-codes {\bf (c,d)} indicate the
cumulative depths corresponding to four considered values of $\kappa'$
in decreasing order from top to bottom.  Each bar-code is split into
ten regions of alternating brightness, representing ten
$10\%$-quantiles of the distribution (e.g., the first dark blue region
of the top bar-code in panel {\bf (c)} indicates that $10\%$ of
reaction events occur till $Dt/R^2 \simeq 1$).}
\label{fig:Ht}
\end{figure*}

Assuming that the domain has a spherical shape and placing the target
at the origin of this domain renders the model exactly solvable.  We
note that although such a geometrical setup is simplified as compared
to realistic situations (e.g., the target site is not necessarily
located at the centre of the domain \cite{Benichou10,Benichou14} or
may be attached to some structure which partially screens it
\cite{Grebenkov17b,Grebenkov2018}), this model captures explicitly two
essential ingredients of the reaction process: the diffusive search
for the target site and its finite intrinsic reactivity.  Importantly,
the fact that the model is exactly solvable, permits us to unveil some
generic features of the full RTD without resorting to any
approximation.

The probability density function $H(r,t)$ of the reaction time $t$ for
a particle released a radial distance $r-\rho$ away from the spherical
target of radius $\rho$ is calculated using standard tools
\cite{Redner,Gardiner,Carslaw}: one first finds the survival
probability $S(r,t)$ of a diffusing particle in a radially symmetric
situation subject to the zero-current boundary condition on the outer
boundary of the domain, and the ``radiation'', or partially-reflecting
boundary condition \cite{Weiss86,Calef83,Shoup82}
\begin{equation}
\label{bc}
D\left.\frac{\partial S(r,t)}{\partial r}\right|_{r=\rho}=\kappa S(\rho,t),
\end{equation}
imposed on the surface of the target site.  The proportionality factor
$\kappa$ in equation \eqref{bc} is an intrinsic rate constant (of
dimension length/time) whose value shows how readily the particle
reacts with the target site upon encounter.  When $\kappa=0$ no
reaction occurs, while the limit $\kappa=\infty$ corresponds to a
perfect reaction, when a particle reacts with the target site upon a
first encounter.  These limiting cases therefore correspond to
perfectly reflecting or absorbing boundaries, respectively.  The RTD
$H(r,t)$ is obtained as the negative derivative of $S(r,t)$ and is
valid for arbitrary values of the system parameters.  Details of these
calculations are presented in the beginning of the Method section.

\begin{figure*}
\includegraphics[width=18cm]{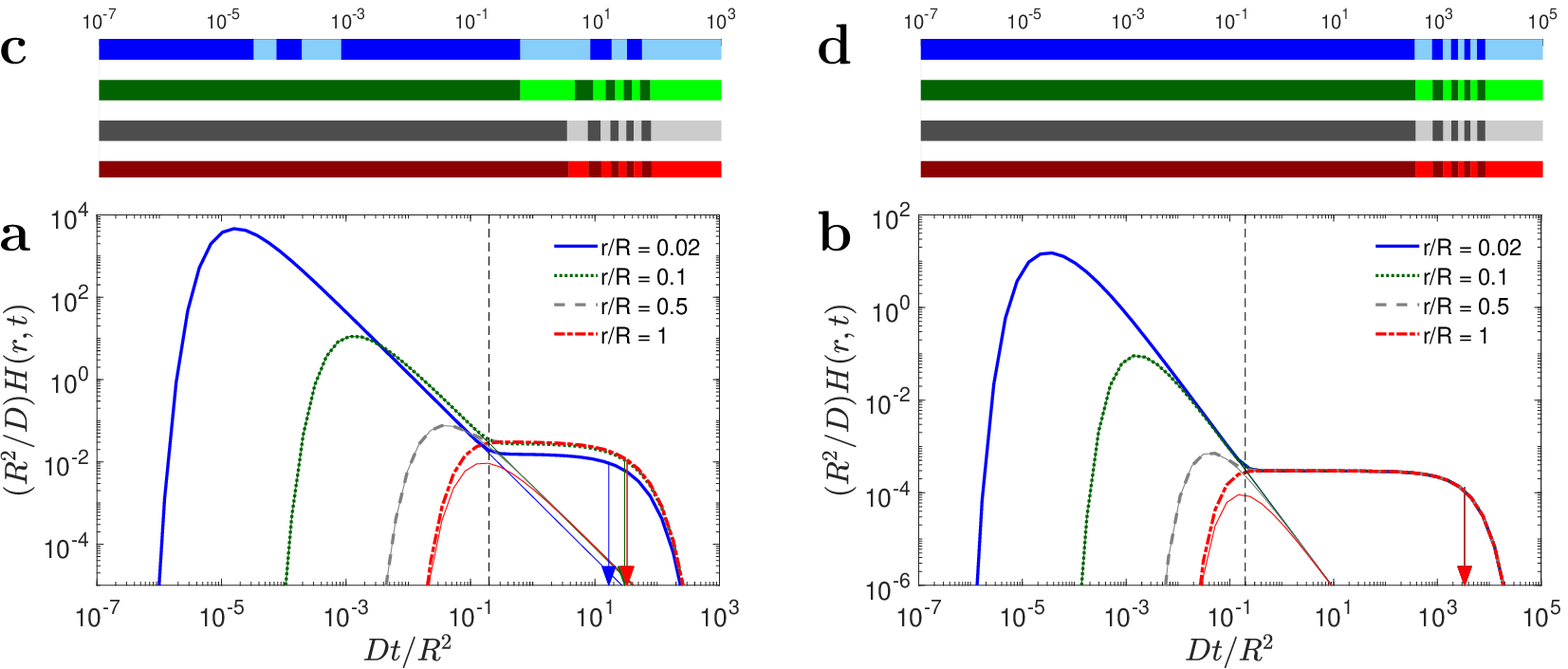}
\caption{
Geometry control.  Reaction time density $H(r,t)$ for a reaction on an
inner target of radius $\rho/R = 0.01$, for the different initial
radii $r$ indicated in the panels ($r$ increasing from top to bottom).
The values of the reactivity are \textbf{(a)} $\kappa = \infty$
(perfectly reactive) and \textbf{(b)} $\kappa R/D =1$ (partially
reactive).  The coloured vertical arrows indicate the mean reaction
times for these cases (note that some arrows coincide).  The vertical
black dashed line indicates the crossover time $t_c =
2(R-\rho)^2/(D\pi^2)$ from the hump-like L\'evy-Smirnov region to a
plateau-like one.  Thin coloured lines show the reaction time density
$H_\infty(r,t)$ from equation \eqref{eq:Ht_Rinf} for the unbounded
case ($R\to\infty$).  The length and time units are fixed by setting
$R=1$ and $R^2/D = 1$.  Clearly the positions of the most likely
reaction times are geometry-controlled by the initial distance to the
target.  Not surprisingly, for the largest initial distance the
solution for the unbounded case underestimates the RTD hump.  Note the
extremely broad range of relevant reaction times (horizontal axis)
spanning over 12 orders of magnitude in panel {\bf (b)}.  Coloured
bar-codes {\bf (c,d)} indicate the cumulative depths corresponding to
four considered values of $r/R$ in increasing order from top to
bottom.  Each bar-code is split into ten regions of alternating
brightness, representing ten $10\%$-quantiles of the distribution.  In
spite of distinctions in the probability densities in panel {\bf (b)},
the corresponding cumulative distributions are close to each other and
result in very similar reaction depths.}
\label{newfig}
\end{figure*}

\begin{figure*}
\includegraphics[width=18cm]{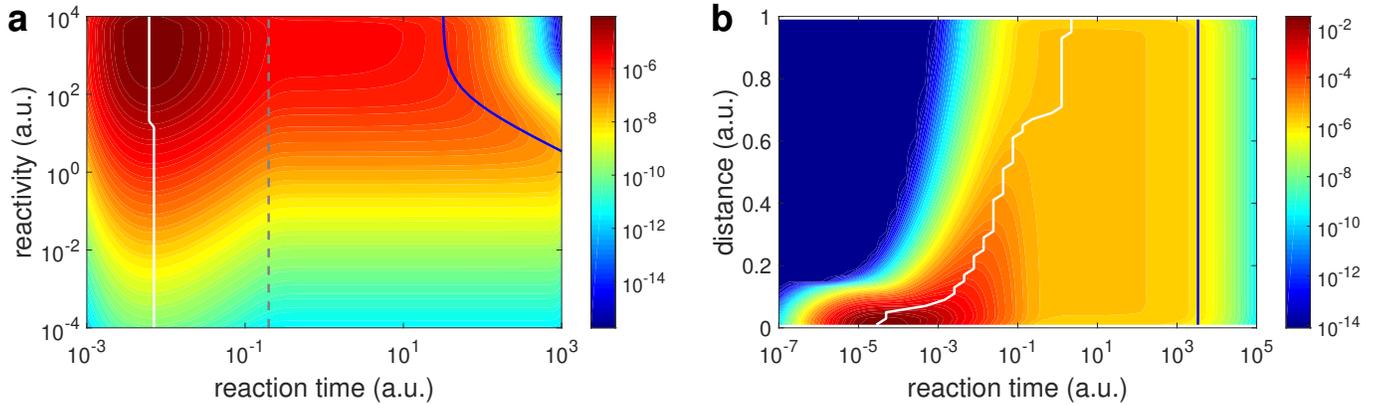} 
\caption{
Reaction versus geometry control.  Impact of the finite reactivity
(reaction control, {\bf a}) and of the distance to the target
(geometry control, {\bf b}) onto the reaction time density shown as a
``heat map'', in which the value of the reaction time density (in
arbitrary units) is determined by the colour code.  Blue and white
lines respectively show the mean and the most probable reaction times
that differ by orders of magnitude.  The grey vertical line indicates
the crossover time $t_c$ that does depend neither on reactivity, nor
on distance.  {\bf (a)} when the reactivity $\kappa$ decreases (with
$r/R = 0.2$ being fixed), the distribution becomes much broader and
extends towards longer reaction times.  {\bf (b)} when the distance to
the target decreases (with $\kappa R/D = 1$ fixed), the most probable
reaction time shifts to the left, whereas the mean reaction time
remains constant.}
\label{fig:Ht_heatmap_kappa}
\end{figure*}

\subsection*{Structure of the full distribution of reaction times}

The typical shapes of the reaction time density $H(r,t)$ are shown in
figure \ref{fig:Ht} for two different release radii $r$ and different
values of the dimensionless reactivity $\kappa R/D$.  Note that the
parameter $\kappa R/D$ represents a combined effect of two factors:
based on the definition of the standard chemical constant
$K_{\mathrm{on}}=4\pi\rho^2\kappa$ for a forward reaction and the
definition of the so-called Smoluchowski constant $K_S=4\pi D\rho$ we
see that $\kappa R/D=(K_{\mathrm{on}}/K_S)(R/\rho)$ and, hence, this
is the ratio of the chemical rate and the Smoluchowski rate constant,
multiplied by the ratio of the sizes of the domain and of the target
site.

We notice that $H(r,t)$ has a much richer structure than the
previously proposed simple form \eqref{benichou}.  The RTD consists of
four distinct time domains seen in figures \ref{fig:Ht}, \ref{newfig},
and \ref{fig:Ht_heatmap_kappa}: first, a sharp exponential cut-off at
short reaction times terminating at the most probable time $t_{\rm
mp}$; second, a region spanning from the most probable reaction time
to the crossover time $t_c$ in which $H(r,t)$ shows a slow power-law
decrease; third, an extended plateau region beyond $t_c$ which
stretches up to the mean reaction time $t_{\rm mean}$; and fourth an
ultimate long-time exponential cut-off.  The shape of the RTD for
varying reactivities highlighting the geometry-controlled
L{\'e}vy-Smirnov hump and the reaction-controlled plateau region is
our central result.  In order to a get a deeper understanding of the
time scales involved in the reaction process, we also introduce and
analyse in the Method section the forms of two complementary
characteristic times: the harmonic mean reaction time $t_{\rm harm} =
1/\langle 1/t\rangle$ and the typical reaction time $t_{\mathrm{typ}}=
t_0 \exp(\langle\ln t/t_0\rangle)$, where the angular brackets denote
averaging with respect to the RTD depicted in figures \ref{fig:Ht} and
\ref{newfig}, and $t_0$ is an arbitrary time scale.  Since the
logarithm is a slowly varying function, its average value is dominated
by the most frequent values of $t$, while anomalously large/small
values corresponding to rare events provide a negligible contribution.
Such an averaged value is widely used to estimate a typical behaviour
in diverse situations \cite{Evans11,Dean14}.

\subsection*{Three characteristic time scales}

The most probable reaction time, corresponding to the very pronounced
maximum, can be calculated explicitly (see the Method section) and has
the approximate form
\begin{equation}
\label{most}
t_{\mathrm{mp}}\approx(r-\rho)^2/(6D).
\end{equation}
Interestingly, this simple estimate, which depends only on the
diffusion coefficient and the initial distance to the target site,
appears to be very robust: $t_{\rm mp}$ indeed shows very little
variation with the reactivity $\kappa$, as one may infer from figures
\ref{fig:Ht} and \ref{fig:Ht_heatmap_kappa}.  In the Method section,
we show that when $\kappa$ decreases from infinity to zero, the value
of $t_{\mathrm{mp}}$ varies only by a factor of $3$.  This
characteristic time is always strongly skewed towards the left tail of
the distribution, that is, to short reaction times: $t_{\mathrm{mp}}$
in fact corresponds to particles moving relatively directly from their
starting point to the target followed by an immediate reaction and
thus generalises the concept of direct, purely geometry-controlled
trajectories \cite{Godec16a} to systems with reaction control.  Note
that expression (\ref{most}) is different from the
diffusion-controlled additive contribution proportional to $1/D$ in
the mean reaction time (\ref{eq:MFPT}).

The second characteristic time scale is the crossover time $t_c$ from the
hump-like L\'evy-Smirnov region specific to an unbounded system, to the
plateau region.  Hence, $t_c$ can be interpreted as the time at which a
molecule starts to feel the confinement.  This can be nicely discerned from
comparison with the density $H_\infty(r,t)$ for the unbounded case (figure
\ref{newfig}).  Thus, reaction times beyond $t_c$ correspond to indirect
trajectories \cite{Godec16a}.  From the result
\begin{equation}
\label{tc}
t_c\approx2(R-\rho)^2/(\pi^2D)
\end{equation}
obtained in the Method section, we see that $t_c$ is independent of
the starting point and of the reactivity $\kappa$, being entirely
dominated by the diffusivity and the difference between the sizes of
the domain $R$ and of the target.  Writing $t_{\rm mp}/t_c =
\pi^2(r-\rho)^2/[12(R-\rho)^2]$, one realises that
the crossover time can be comparable to the most probable time (such
that the hump-like region shrinks), but may also become much larger
than the latter when $r$ is close to $\rho$, as it happens, e.g., when
proteins are produced in a close vicinity of a first gene activated at
$t=0$.  In this case, of course, the hump-like region will be most
pronounced (figure \ref{newfig}).

Finally, the onset of the right exponential shoulder at long reaction
times coincides with the mean reaction time, as indicated by the
arrows in figures \ref{fig:Ht} and \ref{newfig}.  The latter is
obtained from the Laplace transformed distribution (see the Method
section) and is given by the exact formula
\begin{equation}
\label{eq:MFPT}
t_{\mathrm{mean}}=\frac{(r-\rho)(2R^3-\rho r(r+\rho))}{6Dr\rho}+\frac{R^3-
\rho^3}{3\kappa\rho^2},
\end{equation}
which can be thought of as an analogue of the celebrated
Collins-Kimball relation for the apparent reaction rate
\cite{Collins49}.  The first term in equation \eqref{eq:MFPT} is the
standard MFPT to a perfectly reactive target and corresponds to the
classical notion of diffusion-controlled rate.  The additional
contribution to $t_{\mathrm{mean}}$ proportional to $\kappa^{-1}$
accounts for the imperfect reaction with finite reactivity,
independent of the particle's starting point.  When
$t_{\mathrm{mean}}$ is a unique time scale characterising exhaustively
well the reaction kinetics, as it happens for reactions with
sufficiently high concentrations of reactants, one can indeed
distinguish between diffusion or kinetic control.  In
contradistinction, for reactions with nanomolar concentrations of
reactive species, the other time scales $t_{\mathrm{mp}}$ and $t_c$
are equally important and no clear-cut separation between diffusion
and kinetic control can be made.  In the Method section, we also
present an explicit exact expression for the variance of the first
reaction time, which permits us to determine the coefficient of
variation of the RTD and hence, to quantify its broadness.

\subsection*{Geometry versus reaction control}

We emphasise that even for perfect reactions, for which
$\kappa=\infty$, the mean reaction time is orders of magnitude longer
than the most probable reaction time.  For imperfect reactions (finite
$\kappa$ values) the mean reaction time becomes even longer, and
diverges as $1/\kappa$ when $\kappa\to 0$.  The fact that the most
probable reaction time is very weakly dependent on $\kappa$ renders
the difference between the most probable and the mean reaction times
so much more severe for finite $\kappa$.  Another remarkable and so
far unnoticed feature is that a pronounced plateau develops beyond
$t_c$, reflecting an emergent regime of reaction-control.  This
plateau exists even for $\kappa=\infty$ (figure \ref{fig:Ht}) and
becomes increasingly longer with decreasing reactivity $\kappa$,
implying that over several decades the values of the reaction time
become equally probable.  Mathematically speaking this plateau appears
due to the fact that the smallest eigenvalue of the boundary value
problem---the only eigenvalue with an appreciable dependence on
$\kappa$---disentangles from the remaining eigenvalues.  This point is
discussed in more detail in the Method section.  Physically, the
emergence of the plateau implies that the first passage process to the
reaction event becomes even more defocused with decreasing $\kappa$,
i.e., that the spread of possible reaction times increases
significantly.  The long spread of reaction times within this plateau
region is a consequence of geometrically defocused trajectories
exploring the boundary of the reaction volume reinforced by the
necessary multiple collisions with the target before a final reaction
event due to the reaction-control with finite reactivity.  An
important consequence of the existence of the extended plateau region
is that all positive moments of $H(r,t)$, not only the mean reaction
time, will be dominated by integration over this region.  In other
words, the resulting RTD is a concerted effect of geometry-control and
reaction-control.

In figure \ref{newfig} we analyse the effect of the initial distance
to the surface of the target site for both perfect and imperfect
reactions.  The exponential shoulder at long reaction times almost
coincides for all cases, especially when the reactivity is
finite. This part of the reaction time distribution is dominated by
trajectories that equilibrate in the volume before eventual reaction
(indirect trajectories \cite{Godec16a}).  In contrast, we see a strong
variation of the most likely reaction time.  The exponential cut-off
at short reaction times and the position of the maximum of the
distribution is geometry-controlled, as can be anticipated from the
L{\'e}vy-Smirnov form for the unbounded problem (see the Method
section): direct trajectories from the initial position to the target
need a minimum travel time. For increasing initial distance the most
likely reaction time thus moves to longer times and the relative
contribution of the geometry-controlled fraction of direct
trajectories becomes less relevant: instead the particles almost fully
equilibrate in the confined volume until they finally react with the
target.  This reaction-control effect is accentuated for decreasing
reactivity.  We stress that for biological applications both cases are
relevant: shorter initial distances, for instance, are involved when
proteins are produced around a first gene activated at time $t=0$ and
these proteins then need to move to a close-by second gene, here
represented by the inner target.  This scenario is very similar to the
one discussed in reference \cite{Pulkkinen13} as an example for the
rapid search hypothesis \cite{Kolesov07}.  Longer initial distances
are relevant when a molecular signal passes the cellular membrane or
is produced around a cytoplasmic plasmid, and when these molecules
then need to diffuse to the nucleoid region in a bacterial cell or
pass the nuclear membrane in an eukaryotic cell.  Figure
\ref{fig:Ht_heatmap_kappa} summarises the effects of the finite
reactivity and of the distance to the target onto the reaction time
distribution in the form of a ``heat map''.

\subsection*{Short- and long-time behaviour}

We now turn to the discussion of the short- and long time tails of
$H(r,t)$.  The long-time behaviour of the density $H(r,t)$ is
determined by the smallest eigenvalue $\lambda_0$ of the Laplace
operator.  For the spherical domain, one can accurately compute this
eigenvalue by solving the trigonometric equation (see the Method
section).  When both the target and its reactivity are small one gets
$\lambda_0\approx\kappa S_\rho/(DV)$, where the surface area
$S_\rho=4\pi\rho^2$ of the target and the volume of the domain
$V\approx4\pi R^3/3$ are introduced.  According to equation
\eqref{eq:MFPT}, in this limit $t_{\rm mean} \approx1/(D\lambda_0)$, i.e.,
the mean reaction time is dominated by multiple returns to the target
until the reaction occurs.  As the target shrinks ($\rho$ vanishes),
the smallest eigenvalue tends to zero.  In turn, the other eigenvalues
$\lambda_n$, corresponding to rotation-invariant eigenfunctions of the
Laplace operator in the spherical domain, are bounded from below:
$\lambda_n>\pi^2n^2/R^2$ for $n=1,2,\ldots$.  As a consequence, there
is an intermediate range of times, $1/(D\lambda_1)\ll
t\ll1/(D\lambda_0)$, for which the contribution of all higher-order
eigenmodes vanishes, that is, $e^{-Dt\lambda_n}\ll 1$, whereas the
contribution of the lowest eigenmode is almost constant in time,
$e^{-Dt\lambda_0}\approx 1$.  This is precisely the reason why the
intermediate, plateau-like region emerges, see figure
\ref{fig:Ht}.  Note that this region protrudes over an increasing range
of time scales when either the reactivity $\kappa$ or the target
radius $\rho$ decrease, or both.  Note also that this intermediate
regime corresponds approximately to an exponential law which is often
evoked in the context of the first passage statistics to small
targets, see, for instance, references
\cite{Benichou10,Meyer11,Isaacson13}.

While the smallest eigenvalue determines the plateau and the ultimate
exponential cut-off, the short-time behaviour of the reaction time
density $H(r,t)$ is determined by other eigenmodes.  Since the limit
of a small target ($\rho \ll R$) can alternatively be seen as the
limit of large domain size, one can use the density $H_\infty(r,t)$
for diffusion in the exterior of a target, which was first derived by
Collins and Kimball \cite{Collins49},
\begin{eqnarray}
\label{eq:Ht_Rinf}
&& H_\infty(r,t)=\frac{\kappa}{r}\exp\biggl(-\frac{(r-\rho)^2}{4Dt}\biggr)
\biggl\{\frac{\rho}{\sqrt{\pi Dt}}\\
\nonumber
&&-\biggl(1+\frac{\kappa\rho}{D}\biggr)\erfcx\biggl(\frac{r-\rho}{\sqrt{4Dt}}+\biggl(1+\frac{\kappa \rho}{D}\biggr)
\frac{\sqrt{Dt}}{\rho}\biggr)
\biggr\},
\end{eqnarray}
where $\erfcx(x)=e^{x^2} \erfc(x)$ is the scaled complementary error
function (its derivation is reproduced in the Method section).  As
demonstrated in figure \ref{fig:Ht}, equation
\eqref{eq:Ht_Rinf} fully captures the geometry-controlled part of the
reaction time distribution.  In the limit of a perfectly absorbing
target, $\kappa\to\infty$, this expression reduces to
\begin{equation}
\label{eq:Ht_Rinf_kinf}
H_\infty(r,t)=\frac{\rho}{r} \, \frac{r-\rho}{\sqrt{4\pi Dt^3}} \, \exp\biggl(-\frac{(r-\rho)^2}{4Dt}\biggr) ,
\end{equation}
whose normalisation $\rho/r\le1$ reflects the transient nature of
diffusion in three dimensions.  One can easily check that the maximum
of this L{\'e}vy-Smirnov-type density is given exactly by equation
\eqref{most}, as intuitively expected.

\subsection*{Approximate form of the full distribution}

Combining the short and long time contributions we arrive at the
following approximate formula for the reaction time density
\begin{equation}
\label{eq:Ht_app}
H(r,t)\approx H_{\infty}(r,t)+ (1-q)\frac{e^{-t/t_{\rm mean}}}{t_{\rm mean}} \,,
\end{equation}
where $t_{\rm mean}\approx1/(D\lambda_0)$ and 
\begin{equation}
q = \int\limits_0^\infty dt \, H_\infty(r,t) = \frac{\rho/r}{1+D/(\kappa\rho)} < 1 
\end{equation}
is the hitting probability of the target.  The correct normalisation
of $H(r,t)$ is ensured by the prefactor in front of the second term.
Result \eqref{eq:Ht_app} is substantially more general than the simple
form \eqref{benichou} suggested in \cite{Benichou10}.  The form
\eqref{eq:Ht_app} not only extends expression \eqref{benichou} to the
partially-reactive case, i.e., for arbitrary finite values of
$\kappa$, but also emphasises and provides an explicit form for the
contribution from the hump-like region around $t_{\mathrm{mp}}$, which
is most relevant for reactions in which the molecule starts close to
the target.

\begin{figure*}
\includegraphics[width=18cm]{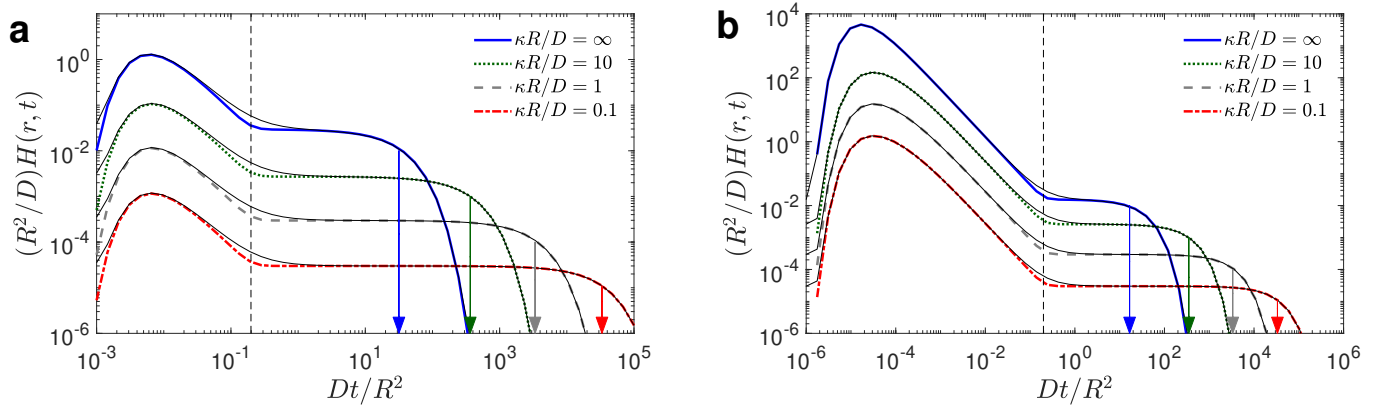}
\caption{
Explicit approximation for the reaction time density $H(r,t)$.  It is
evaluated for a reaction with an inner target of radius $\rho/R=0.01$
with starting point {\bf (a)} $r/R=0.2$ and {\bf (b)} $r/R=0.02$, and
four values of the dimensionless reactivity $\kappa' =
\kappa R/D$ (decreasing from top to bottom).  The coloured vertical
arrows indicate the respective mean reaction time.  The black vertical
dashed line shows the crossover time $t_c=2(R-\rho)^2/(D\pi^2)$ above
which the contribution of higher order Laplacian eigenmodes become
small.  Thin black lines show the approximation \eqref{eq:Ht_app} of
the RTD which very nicely captures the main features of the exact
density.  Length and time units are fixed by setting $R=1$ and
$R^2/D=1$.}
\label{fig:Ht_app}
\end{figure*}

Figure \ref{fig:Ht_app} illustrates the quality of this approximation,
showing that it becomes most accurate when the target radius $\rho$ or
reactivity $\kappa$ are small.  One observes that it accurately
captures both the maximum, the plateau, and the exponential cut-off of
the reaction time distribution.  In turn, the transition between the
maximum and the plateau region is less sharp than in the exact form.
A minor inaccuracy of the approximation \eqref{eq:Ht_app} is that it
reaches a constant---set by the second term---in the short time limit
while the exact distribution vanishes as $t\to0$.  This feature can be
simply removed by multiplying the second term by a Heaviside step
function $\Theta(t-t_c)$ and re-evaluating the normalisation constant.
But even in the present form approximation \eqref{eq:Ht_app} provides
a remarkably good insight into the behaviour of the first passage
dynamics and can thus be used as an efficient and easy-to-handle fit
formula for data analysis or for explicit analytical derivations of
follow-up processes.

\begin{table}
\begin{center}
\begin{tabular}{| c | c | c | c | c | c |}  \hline
$r/R$ & region & $\kappa'=\infty$ & $\kappa'=10$ & $\kappa'=1$ & $\kappa'=0.1$\\
\hline
     & hump-like      & 3.8  & 0.34 & 0.04 & 0.004 \\
0.2  & plateau-like   & 59.4 & 62.9 & 63.2 & 63.2  \\
     & exponential tail      & 36.8 & 36.8 & 36.8 & 36.8  \\  \hline
     & hump-like      & 49.4 & 4.4 & 0.5 & 0.05   \\
0.02 & plateau-like   & 20.0 & 58.8 & 62.7 & 63.15 \\
     & exponential tail      & 30.6 & 36.8 & 36.8 & 36.8 \\   \hline
\end{tabular}
\end{center}
\caption{
Impact of the target reactivity and proximity onto the reaction depth.
Relative weights (in per cents) of three characteristic regions of the
reaction time density for $\rho/R=0.01$: the hump-like region around
the most probable reaction time $t_{\rm mp}$, extending from $0$ till
$t_c = 2(R-\rho)^2/(\pi^2 D)$ (and thus merging two subregions
discussed in the text: the exponential tail left to $t_{\rm mp}$ and
the power-law decay right to $t_{\rm mp}$); the plateau-like region
stretching from $t_c$ to the mean reaction time $t_{\rm mean}$; and
the exponential tail which persists beyond $t=t_{\mathrm{mean}}$.  Two
starting points $r/R$ and four values of dimensionless reactivity
$\kappa'=\kappa R/D$ are used, corresponding to figure \ref{fig:Ht}.}
\label{tab:weights}
\end{table}

\subsection*{Reaction depth}

Lastly, we point out that the contributions of the four different
regimes separated by the time scales $t_{\mathrm{mp}}$, $t_c$, and
$t_{\mathrm{mean}}$ can be further quantified by the corresponding
reaction depths defining which fraction of trajectories reacted up to
a given time.  We thus focus now on the cumulative distribution
function of reaction times
\begin{equation}  
\label{rdepth}
F_r(t)=\int^t_0dt'H(r,t')=1-S(r,t),
\end{equation}
with the evident property $F_r(\infty)=1$ in a bounded domain in which
$H(r,t)$ is normalised, and thus shows explicitly which fraction of
trajectories have reacted up to time $t$.  The reaction depth is
illustrated in the Method section.  Table \ref{tab:weights} summarises
the values of the reaction depths of the three characteristic regions
of the RTD: the hump-like region around $t_{\mathrm{mp}}$, the plateau
region, and the exponential tail.  We realise that for $r/R = 0.2$ the
least amount of the reaction events happens within the hump-like
region: it is of order of just $4\%$ for perfect reactions, and this
fraction rapidly diminishes upon a decrease of $\kappa$.  In turn, a
much larger amount of the reaction events is collected within the
final exponential region.  It is typically of order of almost $37\%$,
independently of the value of $\kappa$, meaning that for such a value
of the ratio $r/R$ roughly one third of all realisations remain
unreacted at time $t = t_{\rm mean}$.  However, most of realisations
of the reaction events occur within the plateau-like regime -- it
amounts to roughly $59\%$ for perfect reactions, and becomes even
bigger for smaller values of $\kappa$.  The situation becomes
different for a smaller release radius: $r/R=0.02$.  Here, for perfect
reactions the majority of trajectories ($49\%$ such that $t_c$ is
close to the median time) react within the hump-like region, while the
plateau region and the final exponential tail contribute only $20\%$
and $30\%$, respectively.  Upon lowering $\kappa$, the hump-like
region is no longer representative, and more reaction events take
place during the exponential tail ($\sim 37\%$) and the plateau-like
regions ($\sim 63\%$), respectively.  In conclusion, the plateau
region appears to be the most important part of the RTD which
contributes most to the overall number of reaction events, except for
the case $r/R \ll 1$ and $\kappa R/D\gg1$, for which the hump-like
region becomes the dominant one.  Concurrently, this plateau is the
region of the strongest defocusing effect, in particular for increased
reaction-control.

\section*{DISCUSSION}
\label{sec:discussion}

Many molecular signalling processes in living biological cells run off
at minute concentrations.  Similarly in vitro experiments tracking the
motion of colloidal particles employ only few particles.  Individual
first passage events in such situations are defocused, that is,
possible reaction times are spread over a vast range comprising orders
of magnitude.  In particular, this implies that any pair of reaction
events will be characterised by highly disparate reaction times.  The
quantitative description of the reaction time to a target in this
scenario therefore cannot simply be based on the mean reaction time.
As we showed, the resulting broad distribution of reaction times is
due to a conspiracy between geometry-control and reaction-control
effects which cannot be disentangled.

We analysed this phenomenon in detail for a generic spherical
geometry, concentrating on several main features.  (i) The reaction
time density consists of four regions with distinct asymptotic
behaviour.  (ii) These time regions are separated by three
characteristic time scales, which means that there is no unique time
scale characterising the kinetic behaviour exhaustively well and the
reaction times are defocused.  In consequence, the textbook notions of
diffusion versus kinetic control, which are appropriate for reactions
operating at abundant concentrations, are not applicable in our case.
We explicitly determined these times scales and also the associated
reaction depths.  (iii) A finite reactivity broadens an intermediate
regime characterised by an extended plateau region.  We showed that
the plateau emerges due to a time scale separation of the lowest and
the next eigenvalues of the diffusion-controlling Laplace operator.
The fundamental parameter we found to quantify this intermediate
regime is the reaction-control represented by the dimensionless
reactivity $\kappa R/D$.  A majority of the reaction events occur
within this region, except for the case $r/R \ll 1$ and $\kappa R/D
\gg 1$.  In turn, for perfect reactions with a reactant starting very
close to the target site the most important part of the RTD is the
hump-like region which contributes with almost $50$ per cent of the
reaction events.  (iv) The geometry control of the initial
particle-to-target distance strongly affects the position and the
amplitude of the maximum of the reaction time distribution and thus
the most likely reaction time.  (v) We came up with a simple and thus
practical approximative formula for the full reaction time
distribution.  In particular, we demonstrated that this approximation
captures both the most likely and mean reaction times.  While the
derivation relied on the rotation symmetry of the considered geometric
domain, this approximation is expected to be valid in more complex
confinements, as long as the target site is far enough from the
surrounding outer boundary.  Our main conclusion is that
reaction-control with finite reactivity leads to even stronger
reaction time defocusing, stressing the necessity to know the full
RTD.  This conclusion will serve as a benchmark for the behaviour in
geometrically more complex situations \cite{Benichou14} when, e.g.,
the target site is on the wall or bound to some geometrical structure
within the domain, and a fully analytical solution is impossible.

\section*{METHOD}
\subsection*{Exact distribution of reaction times}

We consider a diffusion process in a three-dimensional domain $\Omega=
\{\x\in\R^3:\rho<\|\x\|<R\}$ between two concentric spheres -- a small
target and a bounding surface of radii $\rho$ and $R$, respectively.
Although the solution of the underlying diffusion problem is well
known \cite{Redner,Carslaw}, we rederive it here for completeness and
to highlight several practical points discussed in the main text.  In
fact, the Laplace transformed probability density function $\tilde
H(\x,p)$ satisfies the modified Helmholtz equation
\begin{equation}
\label{eq:Helmholtz}
(p-D\Delta)\tilde{H}(\x,p)=0\qquad(\x\in\Omega),
\end{equation}
subject to the boundary conditions
\begin{subequations}
\begin{eqnarray}
&&\bigl(\partial_n\tilde{H}(\x,p)\bigr)|_{\|\x\|=R}=0,\\[0.2cm]
&&\biggl(\frac{D}{\kappa}\partial_n\tilde{H}(\x,p)+\tilde{H}(\x,p)\biggr)|_{\|\x\|
=\rho}=1.
\end{eqnarray}
\end{subequations}
Here $\Delta$ is the Laplace operator, $D$ is the diffusion
coefficient, $\kappa$ is the intrinsic reactivity, and $\partial_n$ is
the normal derivative directed outward from the domain $\Omega$.

The rotational symmetry of the domain reduces the partial differential
equation \eqref{eq:Helmholtz} to an ordinary differential equation
with respect to the radial coordinate $r$,
\begin{subequations}
\begin{eqnarray}
&&\tilde{H}''(r,p)+\frac{2}{r}\tilde{H}'(r,p)-\frac{p}{D}\tilde{H}(r,p)=0,\\[0.2cm]
&&\tilde{H}'(R,p)=0,\\[0.2cm]
&&\biggl(\tilde{H}(r,p)-\frac{D}{\kappa}\tilde{H}'(r,p)\biggr)_{r=\rho}=1,
\end{eqnarray}
\end{subequations}
where primes denote derivatives with respect to $r$.  The solution of
this equation is
\begin{equation}
\label{eq:Hp}
\tilde{H}(r,p)=\frac{g(r)}{g(\rho)-g'(\rho)\frac{D}{\kappa}} \,,
\end{equation}
where
\begin{equation}
\label{eq:gr}
g(r)=\frac{R\sqrt{p/D}\cosh\xi-\sinh\xi}{r\sqrt{p/D}} \,,
\end{equation}
with $\xi=(R-r)\sqrt{p/D}$. It follows that
\begin{equation}
\label{eq:dgr}
g'(r)=\frac{(1-Rr p/D)\sinh\xi-\xi\cosh\xi}{r^2\sqrt{p/D}} \,.
\end{equation}
The mean reaction time is obtained from the Laplace-transformed
density as
\begin{equation}
t_{\rm mean} = -\lim\limits_{p\to 0}\frac{\partial}{\partial p}\tilde H(r,p) \,,
\end{equation}
from which equation \eqref{eq:MFPT} follows.  

In the limit $R\to \infty$, equations \eqref{eq:Hp}, \eqref{eq:gr},
and \eqref{eq:dgr} yield
\begin{equation}
\label{eq:Htilde_inf}
\tilde{H}_\infty(r,p)=\frac{(\rho/r) \, e^{-(r-\rho)\sqrt{p/D}}}{1+\bigl(1+ \rho \sqrt{p/D}\bigr) D/(\kappa \rho)} \,.
\end{equation}
Due to the transient character of three-dimensional diffusion, the
related distribution is not normalised to unity, but
$\tilde{H}_\infty(r,p=0)=(\rho/r)/(1 +D/(\kappa\rho))<1$ is the
probability of reacting with the target before escaping to infinity.
The inverse Laplace transform yields equation \eqref{eq:Ht_Rinf}.
Using the relation $\tilde{S}_\infty(r,p) = (1 -
\tilde{H}_\infty(r,p))/p$ and equation \eqref{eq:Htilde_inf}, one can
also compute the survival probability $S_\infty(r,t)$ in the time
domain
\begin{eqnarray}
\nonumber
&& S_\infty(r,t) = 1 - \frac{\rho \exp\bigl(-\frac{(r-\rho)^2}{4Dt}\bigr)}{r(1 + D/(\kappa\rho))} 
\biggl\{ \erfcx\biggl(\frac{r-\rho}{\sqrt{4Dt}}\biggr) \\
\label{eq:S_inf}
&& - \erfcx\biggl(\frac{r-\rho}{\sqrt{4Dt}} + \left(1 + \frac{\kappa \rho}{D}\right) \frac{\sqrt{Dt}}{\rho}\biggr)\biggr\} \,.
\end{eqnarray}

The Laplace inversion of equation \eqref{eq:Hp} can be performed by
identifying the poles of the function $\tilde{H}(r,p)$ in the complex
plane $p\in\C$, that is, by finding the zeros of the function
\begin{equation}
F(p)=g(\rho)-\frac{D}{\kappa}g'(\rho).  
\end{equation}
For convenience, we introduce dimensionless Laplace variable
$s=(R-\rho)^2p/D$, so that
\begin{align}
\nonumber
&& F(p) = \frac{1}{\rho^2\sqrt{s}}\biggl(\bigl(\rho R+\mu(R-\rho)^2\bigr)\sqrt{s}\cosh
\sqrt{s}\\
&&-\bigl(\rho(R-\rho)+\mu(R-\rho)^2-\mu R\rho s\bigr)\sinh\sqrt{s}\biggr),
\end{align}
where we defined the dimensionless ``dilatoriness'' parameter $\mu$ as
\begin{equation}
\label{mu}
\mu=\frac{D}{\kappa(R-\rho)} \,.  
\end{equation}
The perfectly reactive target with $\kappa=\infty$ corresponds to
$\mu=0$.  In other words, for high reactivity $\kappa$ the value of
the dilatoriness $\mu$ is small and reactions occur more likely on
first encounter, and vice versa.  Note that a fully reflecting target
with $\kappa=0$ is excluded from our analysis because the reaction
time would be infinite. In other words, we always consider
$0\leq\mu<\infty$.

The solutions of the equation $F(p)=0$ lie on the negative real axis.
Setting $s=- \alpha^2$, one gets the trigonometric equation
\begin{equation}
\label{eq:eq_alpha}
\tan\alpha=\frac{\alpha\bigl(\rho R+\mu(R-\rho)^2\bigr)}{\rho(R-\rho)+\mu(R-
\rho)^2+\mu R\rho\alpha^2} \,.
\end{equation}
This equation has infinitely many positive solutions that we denote as
$\alpha_n$, with $n=0,1,2,\ldots$ Since the function on the right-hand
side has the slope $\frac{\rho
R+\mu(R-\rho)^2}{\rho(R-\rho)+\mu(R-\rho)^2}>1$ near $\alpha=0$, the
smallest solution $\alpha_0$ lies in the interval $(0,\pi/2)$.  More
generally, the $n$th solution lies in the interval $(\pi
n,\pi(n+1/2))$ and tends, for any fixed $\kappa$, to the left boundary
of the interval as $n \to\infty$.  Note that $\alpha =0$ (or $p=0$) is
not a pole of the function $\tilde H(r,p)$.

Once the poles are identified, we determine the residues by taking the
derivative of $F(p)$ at the poles.  Applying the theorem of residues
to compute the inverse Laplace transform, we finally deduce the exact
expression for the probability density $H(r,t)$ of the reaction time
for a particle starting at a distance $r - \rho$ from the target,
\begin{equation}  \label{eq:Ht}
H(r,t) = \sum\limits_{n=0}^\infty u_n(r) \, e^{- Dt \lambda_n} ,
\end{equation}
with
\begin{eqnarray}
\lambda_n &=& \alpha_n^2/(R-\rho)^2 , \\  \label{eq:un}
u_n(r) &=& c_n \, \frac{D}{(R-\rho)^2} \\  \nonumber
&\times& \frac{R\alpha_n\cos\bigl(\alpha_n\frac{R-r}{R-\rho}\bigr)-(R-\rho)\sin
\bigl(\alpha_n\frac{R-r}{R-\rho}\bigr)}{r\alpha_n},
\end{eqnarray}
where the expansion coefficients $c_n$ are given explicitly by the
residues as
\begin{equation}
\label{cn}
c_n=\frac{2\rho^2\alpha_n^2}{(\rho R+\mu(R^2+\rho^2))\alpha_n\sin\alpha_n+\rho(
\mu R\alpha_n^2-\rho)\cos\alpha_n} \,.
\end{equation}

\subsection*{Long-time behavior of the RTD}

When either the target radius $\rho$ is small or the dilatoriness
parameter $\mu$ is large, the slope of the right-hand side of equation
\eqref{eq:eq_alpha} is close to unity and thus the smallest eigenvalue
$\alpha_0$ is close to zero.  Expanding both sides of equation
\eqref{eq:eq_alpha} into Taylor series one finds the first-order
approximation
\begin{eqnarray}
\alpha_0&\simeq&\frac{\rho}{\sqrt{\rho(R-\rho)+\mu(R-\rho)^2}}\\  \nonumber
&\times&\biggl(\frac13+\mu R\rho\frac{\rho R+\mu(R-\rho)^2}{(\rho(R-\rho)+
\mu(R-\rho)^2)^2}\biggr)^{-1/2}+\ldots.
\end{eqnarray}
In particular, for small target radius, $\rho\to0$, at fixed
dilatoriness $\mu$ we see that
$\alpha_0\simeq\sqrt{3}(\rho/R)\mu^{-1/2}$.  In turn, when $\mu\to
\infty$ with fixed $\rho$,
\begin{equation}
\alpha_0\simeq\frac{\sqrt{3}\rho}{\sqrt{R^2+R\rho+\rho^2}} \,\mu^{-1/2}.
\end{equation}
In both cases $\alpha_0$ is proportional to $\rho$ and inversely
proportional to $\sqrt{\mu}$.  As a consequence, the term with the
slowest decay time behaves as
\begin{equation}
\begin{split}
\lambda_0 &\simeq \frac{3\kappa \rho^2}{D(R-\rho)(R^2+R\rho+\rho^2)}
\simeq \frac{3\kappa \rho^2}{D R^3} \approx \frac{\kappa S_\rho}{DV} \,,
\end{split}
\end{equation}
where in the intermediate approximation we ignored terms of order
$\rho/R$ and higher, and we introduced the surface area
$S_\rho=4\pi\rho^2$ of the target and the volume of the domain
$V\approx4\pi R^3/3$.

We also note that the approximation
$c_0\approx3(\rho/R)^2/(\mu+3\rho/(2R))$ holds for $\rho\ll R$, and
thus $c_0/\alpha_0^2\simeq1/(1+3\rho/(2\mu R))$, i.e., it is close to
unity as long as the dilatoriness $\mu$ is not too small.  Therefore
the survival probability can be accurately approximated as $S(r,t)
\simeq\exp(-Dt\alpha_0^2/R^2)$ for intermediate and large times.  In
this case the median reaction time becomes
\begin{equation}  
t_{\rm median}\approx\frac{R^2\ln2}{D\alpha_0^2}\simeq\frac{\mu R^4\ln 2}{3D\rho^2}\simeq
\frac{R^3\ln 2}{3\kappa\rho^2} ,
\end{equation}
from which the relation $t_{\rm median}\approx t_{\rm mean}\ln2$.
This median value is close to the mean reaction time which in the
limit $\rho\ll R$ has the dominant behaviour as $R^3/(3\kappa\rho^2)$
according to equation \eqref{eq:MFPT}. In turn, the most probable
reaction time, which is determined by the higher-order eigenmodes, is
orders of magnitude smaller.  This behaviour is, however, only present
for weakly reactive targets.  In contrast, the median time for perfect
reactions is usually close to the crossover time $t_c$, while
$t_{\mathrm{mean}}$ is orders of magnitude larger.

\subsection*{Most probable reaction time}

\begin{figure}
\begin{center}
\includegraphics[width=8.8cm]{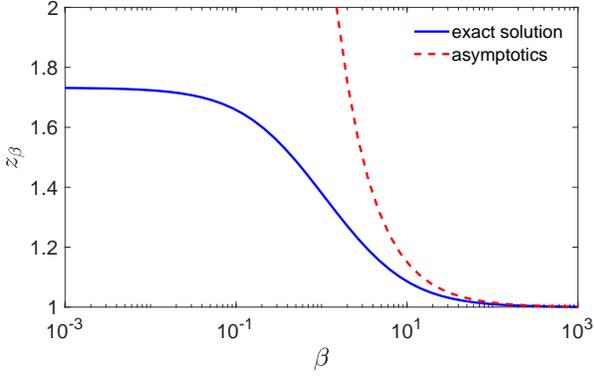} 
\end{center}
\caption{
Weak dependence of the most probable reaction time on reactivity.  The
numerical solution $z_\beta$ of equation
\eqref{eq:tmp_eq} as a function of $\beta$ (solid line) and its
large-$\beta$ asymptotic behavior \eqref{eq:zbeta_large} shown by
dashed line. }
\label{fig:zbeta}
\end{figure}

One may deduce from figure \ref{fig:Ht} that the region around the
most probable reaction time is well described by the function in
\eqref{eq:Ht_Rinf}, which corresponds to the solution in the limit $R
\to \infty$.  Hence, the most probable reaction time $t_{\rm mp}$ can be
obtained with a good accuracy by merely differentiating this function
with respect to $t$ and setting the result equal to zero: 
\begin{equation}  
\label{tt}
t_{\rm mp} = \frac{(r - \rho)^2}{6 D} z^2 \,,
\end{equation}
where $z$ is defined implicitly as the solution of the following,
rather complicated transcendental equation
\begin{align}  
& \beta^2 z^4 - 3(1 + \beta) z^2 + 9 \nonumber \\
& - \sqrt{\pi/6}\, \beta^3 z^5 \erfcx\biggl(\frac{\sqrt{3/2}}{z} + \frac{\beta z}{\sqrt{6}}\biggr) = 0,  \label{eq:tmp_eq}
\end{align}
where ${\rm erfcx(x)}$ is the scaled complementary error function, and
\begin{equation}    \label{eq:beta}
\beta = \frac{r-\rho}{\rho} \biggl(1 + \frac{\kappa \rho}{D} \biggr) \,.
\end{equation}
We denote the solution of this equation as $z_\beta$.  When $\beta$
tends to $0$, a Taylor expansion of the left-hand side of
\eqref{eq:tmp_eq} yields $z^2 - 9 + O(\beta)$, from which $z_0 =
\sqrt{3}$.  In the opposite limit $\beta\to\infty$, one uses the
asymptotic behaviour of the function $\erfcx(x)$ to get

\begin{equation} \label{eq:zbeta_large}
z_\beta \simeq 1 + \frac{3}{2\beta} + O(\beta^{-2}) .
\end{equation}
With some technical efforts, one can prove that $z_\beta$ is a
monotonously decreasing function of $\beta$ (see
Fig. \ref{fig:zbeta}).  We conclude that $z_\beta$ is bounded between
$\sqrt{3}$ and $1$ so that the most probable time $t_{\rm mp}$ lies
between $(r - \rho)^2/(6D)$ (for $\kappa \rho \gg 1$) and
$(r-\rho)^2/(2D)$ (for $\kappa \rho \ll 1$).  In other words, the most
probable reaction time shows remarkably weak dependence on the
reactivity $\kappa$, as illustrated by Fig. \ref{fig:zbeta}.

\subsection*{Moments of the reaction time}

As we have already remarked in the main text, the positive moments of
the RTD of an arbitrary order are dominated by the integration over
the plateau-like region such that their values appear close to the
onset of the crossover to the final region -- the exponential decay of
the RTD.  The exact values of the positive moments of the random
reaction time $\tau$ can be accessed directly by a mere
differentiation of $\tilde{H}(r,p)$ with respect to the Laplace
parameter $p$ and subsequently taking the limit $p = 0$:
\begin{equation}
\langle \tau^k \rangle = (-1)^k \lim\limits_{p\to 0} \frac{\partial^k \tilde{H}(r,p)}{\partial p^k} \,.
\end{equation}
For instance, a lengthy but straightforward calculation yields
the exact formula for the variance of the reaction time:
\begin{align}   \label{eq:variance}
& \langle \tau^2 \rangle - \langle \tau \rangle^2 = \frac{1}{90 D^2 r^2 \rho^4} \biggl\{ 
10r^2(R^3-\rho^3)^2  (D/\kappa)^2 \\ \nonumber
&+ 4\rho r^2 (5R^3 + 6R^2\rho + 3R\rho^2 + \rho^3)(R-\rho)^3 (D/\kappa) \\  \nonumber
&+ \rho^2(r-\rho)\bigl(2R^3(5R^3\rho + 5R^3 r + 10 r^2\rho^2 - 18R^2 r\rho) \\  \nonumber
& - \rho^2 r^2(\rho+r)(r^2+\rho^2) \bigr)\biggr\} ,
\end{align}
from which one also gets the coefficient of variation, $\gamma =
\sqrt{\langle \tau^2 \rangle - \langle \tau \rangle^2}/\langle
\tau\rangle$, which characterises fluctuations of the random
reaction time $\tau$ around its mean, i.e., the effective broadness of
the reaction time density.  As compared to Ref. \cite{Moffitt10}, the
expressions \eqref{eq:MFPT} and \eqref{eq:variance} permit us to
quantify the effect of both rate-controlling factors.  

For a perfectly reactive target, the coefficient of variation diverges
as the starting point $r$ approaches $\rho$, in particular, one gets
\begin{equation}  \label{eq:gamma_perfect}
\gamma^2 \simeq \frac{2\rho}{r - \rho} + O(1),
\end{equation}
when the target is small or the confining domain is large ($\rho \ll
R$).  In turn, for a partially reactive target, the squared
coefficient of variation is finite in the limit $r\to\rho$ and for a
small target reads
\begin{equation}  \label{eq:gamma_partial}
\gamma^2 \simeq 1 + \frac{2\rho \kappa}{D} \,.
\end{equation}
The coefficient of variation $\gamma$ in equations
\eqref{eq:gamma_perfect} and \eqref{eq:gamma_partial} exceeds $1$,
allowing one to classify this distribution as broad, according to the
standard terminology in statistics \cite{Mejia11,Mattos12,Mattos14}.
In both cases, the asymptotic behaviour of $\gamma$ does not depend on
the size of the confining domain, $R$.

We turn next to the negative order moments of the RTD which are
clearly dominated by the region close to the origin and hence, probe
the left tail of the distribution.  The computation of negative
moments (with $\nu > 0$) involves integration:
\begin{equation}  \label{eq:tmean_harm0}
\langle \tau^{-\nu} \rangle = \int\limits_0^\infty dt \, t^{-\nu} \, H(r,t) 
= \frac{1}{\Gamma(\nu)} \int\limits_0^\infty dp \, p^{\nu-1} \, \tilde{H}(r,p) .
\end{equation}
Although this integral is expressed in terms of the explicitly known
Laplace transform $\tilde{H}(r,p)$ from equation \eqref{eq:Hp}, its
analytical evaluation does not seem to be feasible.

In turn, the integral takes a more tractable form in the limit
$R\to\infty$ corresponding to diffusion in the exterior of a partially
reactive target of radius $\rho$.  Due to the transient character of
diffusion in three dimensions, the probability density $H_\infty(r,t)$
is not normalised to $1$ as the molecule can escape to infinity.  The
integral of the density $H(r,t)$ yields thus the probability of
reacting at the target:
\begin{equation} \label{eq:q}
q = \tilde{H}_\infty(r,p=0) = \frac{\rho/r}{1 + D/(\kappa \rho)} \,.
\end{equation}
The negative order moments of the renormalised density
$H_\infty(r,t)/q$ are
\begin{equation}  \label{eq:tau_mean}
\langle \tau^{-\nu} \rangle_n = \frac{2}{\Gamma(\nu)} \biggl(\frac{D}{(r-\rho)^2}\biggr)^\nu  
 \int\limits_0^\infty dz \, \frac{z^{2\nu-1} e^{-z}}{1+ z/\beta} \,,
\end{equation}
where $\beta$ was defined in \eqref{eq:beta}.  In the limit $\kappa
\to \infty$, one finds
\begin{equation}  \label{eq:tau_mean_kinf}
\langle \tau^{-\nu} \rangle_n = \frac{2}{\Gamma(\nu)} \biggl(\frac{D}{(r-\rho)^2}\biggr)^\nu \Gamma(2\nu) .
\end{equation}

While the mean reaction time diverges for the exterior problem, the
negative order moments are well defined and can thus characterise the
reaction process.  In particular, the harmonic mean reaction time,
defined as
\begin{equation}
t_{\rm harm} = \frac{1}{\langle \tau^{-1} \rangle_n} \,, 
\end{equation}
is deduced from \eqref{eq:tau_mean} for $\nu = 1$:
\begin{equation}  \label{eq:tharm}
t_{\rm harm} = \frac{(r-\rho)^2}{2D} \, \beta^{-1}\biggl(1 - \beta e^{\beta} \Ei_1(\beta)\biggr)^{-1} \,,
\end{equation}
where $\Ei_1(z) = \int\nolimits_1^\infty dx \,e^{-zx}/x$ is the
exponential integral.  The dependence of the harmonic mean on the
reactivity $\kappa$ is fully captured via $\beta$.  In the limit
$\kappa \to \infty$, this mean approaches
\begin{equation}  \label{eq:tharm_kinf}
t_{\rm harm} = \frac{(r-\rho)^2}{2D} \,,
\end{equation}
and is thus of the order of the most probable time, representing the
relevant time scale of the problem.  In the opposite limit $\kappa \to
0$, $\beta$ approaches a constant, and the harmonic mean reaction time
also reaches a constant.  One can check that $t_{\rm harm}$
monotonously decreases as $\beta$ (or $\kappa$) grows.

\begin{figure*} 
\begin{center}
\includegraphics[width=18cm]{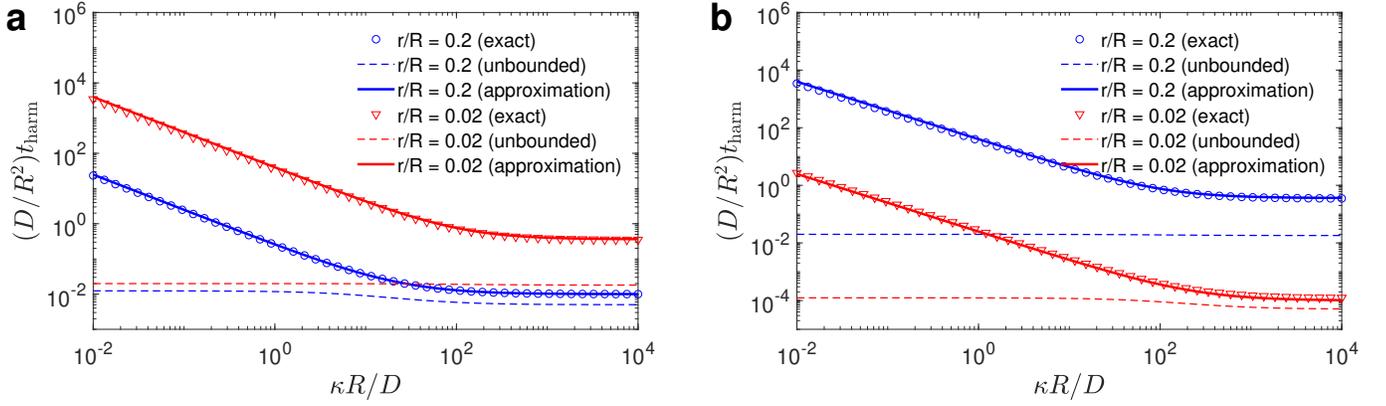} 
\end{center}
\caption{
The harmonic mean reaction time, $t_{\rm harm}$, as a function of the
dimensionless reactivity, $\kappa' = \kappa R/D$.  {\bf (a)} An inner
target has a radius $\rho/R = 0.1$ (blue curves) or $\rho/R = 0.01$
(red curves), and the release radius $r/R = 0.2$.  Symbols show the
results for the concentric domain, obtained by a numerical evaluation
of the integral in equation \eqref{eq:tmean_harm0} with $\nu = 1$;
dashed lines present the relation \eqref{eq:tharm} for the exterior
problem; solid lines indicate the relation \eqref{eq:tharm} multiplied
by the reacting probability $q$ from \eqref{eq:q}.  The length and
time units are fixed by setting $R=1$ and $R^2/D = 1$. {\bf (b)} The
same but for $\rho/R = 0.01$ and two values of $r/R$: $0.2$ (blue
curves) and $0.02$ (red curves). }
\label{fig:tharm}
\end{figure*}

Figure \ref{fig:tharm} illustrates by dashed lines the behaviour of
the function in \eqref{eq:tharm}, in particular, its approach to the
limiting expression \eqref{eq:tharm_kinf} as $\kappa$ increases.  One
can appreciate a very weak dependence of the harmonic mean reaction
time for the exterior problem on the reactivity $\kappa$.  We also
show the harmonic mean reaction time in the concentric domain,
obtained by a numerical integration in equation \eqref{eq:tmean_harm0}
with $\nu = 1$.  This mean significantly depends on $\kappa$ and
behaves as $1/\kappa$ for small $\kappa$.  Given that the probability
density $H(r,t)$ for the concentric domain can be accurately
approximated by $H_\infty(r,t)$ at small times (see equation
\eqref{eq:Ht_app}), the harmonic mean reaction time for the concentric
domain can be approximated by the expression in
\eqref{eq:tmean_harm0}, multiplied by the reaction probability $q$.
This approximation, shown by solid lines, turns out to be remarkably
accurate when the target radius $\rho$ is small as compared to the
radius $R$ of the confining domain.  We can also conclude that the
significant variations of $t_{\rm harm}$ with $\kappa$ for the
concentric domain come from those of $q$ with $\kappa$.

Finally, we consider the time scale
\begin{equation}
t_{\mathrm{typ}} = t_0 \exp(\langle \ln(t/t_0) \rangle)
\end{equation}
(where $t_0$ is an arbitrary time scale), based on the mean logarithm
of the reaction time -- an important characteristic of the reaction
process, which emphasizes the typical values of $t$, i.e., values
observed in most of experiments.  Indeed, the logarithm is a
slowly-varying function and its average is supported by the most
frequently encountered values of $t$ with the rare anomalously long or
short reaction times being effectively filtered out.  The estimates
based on $t_{\mathrm{typ}}$ are widely used in the analysis of
stochastic reaction-diffusion or transport process in random
environments (see, e.g., Refs. \cite{Evans11,Dean14} and references
therein).  Such an averaged value can be formally computed as
\begin{eqnarray}
&& \langle \ln(\tau/t_0) \rangle = \sum\limits_{n=0}^\infty u_n(r) \int\limits_0^\infty dt \, \ln(t/t_0) e^{-Dt\lambda_n} \\  \nonumber
&&~~~= -\sum\limits_{n=0}^\infty u_n(r) \frac{\gamma + \ln(Dt_0\lambda_n)}{D\lambda_n} \\  \nonumber
&&~~~= \biggl(\ln \frac{(R-\rho)^2}{Dt_0}\biggr) - \gamma - \frac{(R-\rho)^2}{D} \sum\limits_{n=0}^\infty u_n(r) \frac{\ln \alpha_n^2}{\alpha_n^2} \,,
\end{eqnarray}
where $\gamma \approx 0.5772\ldots$ is the Euler constant, from which
\begin{equation}  \label{eq:tlog}
t_{\rm typ} = \frac{(R-\rho)^2}{D} \exp\biggl(- \gamma - \frac{(R-\rho)^2}{D} \sum\limits_{n=0}^\infty u_n(r) \frac{\ln \alpha_n^2}{\alpha_n^2} \biggr),
\end{equation}
where $u_n(r)$ are given by \eqref{eq:un}.

To get a more explicit dependence on the initial radius $r$, one can
again consider the exterior problem ($R = \infty$).  Rewriting
equation \eqref{eq:tau_mean} as
\begin{eqnarray} 
\langle \tau^{-\nu} \rangle_n &=& \biggl(\frac{D}{(r-\rho)^2}\biggr)^\nu  \frac{2\Gamma(2\nu)}{\Gamma(\nu)} \\  \nonumber
&\times& \biggl(1 - \frac{1}{\beta \Gamma(2\nu)}  \int\limits_0^\infty dz \, \frac{z^{2\nu} e^{-z}}{1+ z/\beta} \biggr) \,,
\end{eqnarray}
in order to get a Taylor expansion as $\nu \to 0$, one gets
\begin{equation}
\langle \ln (\tau/t_0) \rangle_n = 
\biggl\{\ln \biggl(\frac{(r-\rho)^2}{D t_0}\biggr) + \gamma + 2e^{\beta} \Ei_1(\beta)\biggr\},
\end{equation}
where the expectation is computed with respect to the renormalised
density $H_\infty(r,t)/q$.  We obtain thus the logarithmic mean time
\begin{equation}  \label{eq:Tlog}
t_{\rm typ} = \frac{(r-\rho)^2}{D}  \exp\bigl(\gamma + 2e^{\beta} \Ei_1(\beta)\bigr) \,.
\end{equation}
In the limit $\kappa\to\infty$, $e^{\beta} \Ei_1(\beta)$ vanishes as
$1/\beta$, so that for a perfectly reactive target one gets
\begin{equation}  \label{eq:Tlog_kinf}
t_{\rm typ} = \frac{(r-\rho)^2}{D} \, e^{\gamma},
\end{equation}
which signifies that in the limit $\kappa = \infty$ the logarithmic
mean time is comparable to the most probable reaction time $t_{\rm
mp}$.

\begin{figure} 
\begin{center}
\includegraphics[width=8.8cm]{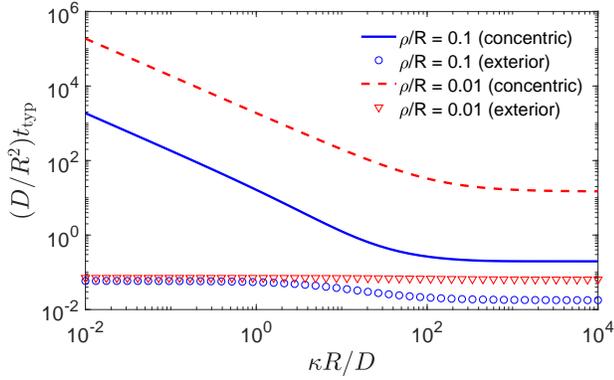} 
\end{center}
\caption{
The logarithmic mean reaction time, $t_{\rm typ}$, as a function of
the dimensionless reactivity, $\kappa R/D$.  $t_{\rm typ}$ is
evaluated for an inner target of radius $\rho/R = 0.1$ (blue curves)
or $\rho/R = 0.01$ (red curves), for the initial radius $r/R = 0.2$.
Lines show the results for the concentric domain from equation
\eqref{eq:tlog}, whereas symbols present the relation \eqref{eq:Tlog}
for the exterior problem.  The length and time units are fixed by
setting $R=1$ and $R^2/D = 1$. }
\label{fig:tlog}
\end{figure}

Figure \ref{fig:tlog} shows the logarithmic mean reaction time,
$t_{\rm typ}$, as a function of the dimensionless reactivity $\kappa
R/D$.  As for the harmonic mean in Fig. \ref{fig:tharm}, the results
for a bounded concentric domain ($R = 1$) and for the exterior problem
($R = \infty$) differ significantly.  The particular definition of the
logarithmic time does not allow one to easily renormalise $t_{\rm
typ}$ for the exterior domain to get an approximation for the bounded
domain.

\begin{figure*}  
\begin{center}
\includegraphics[width=18cm]{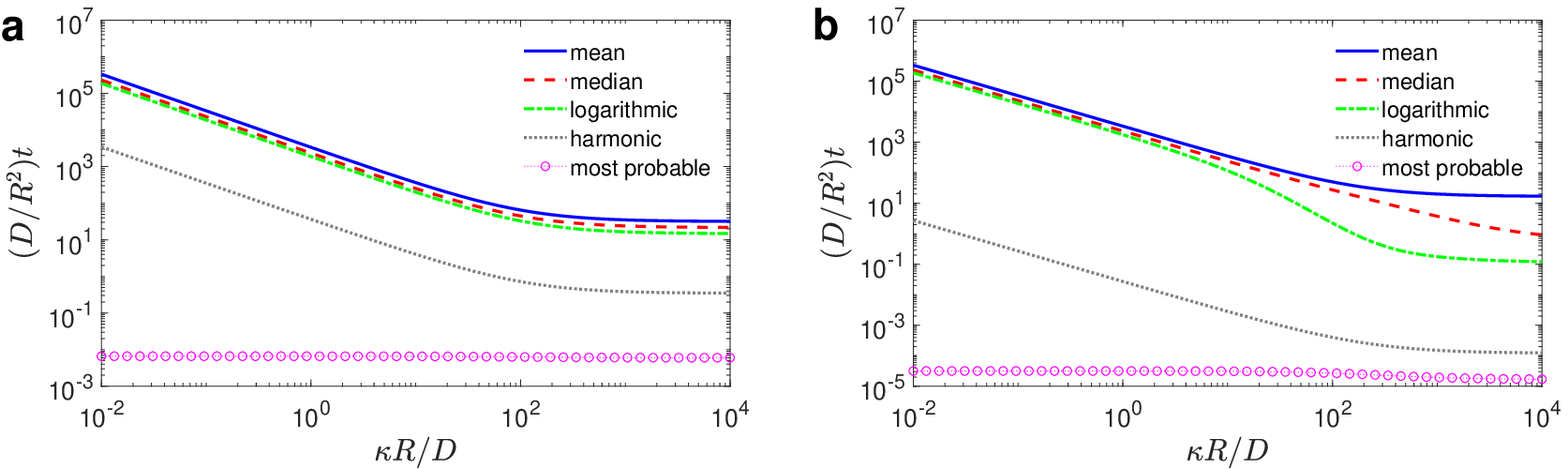}
\end{center}
\caption{
Comparison of several means of the reaction time.  They are evaluated
as functions of the normalised reactivity, $\kappa R/D$, for an inner
target of radius $\rho/R = 0.01$ and the starting point $r/R = 0.2$
{\bf (a)} and $r/R = 0.02$ {\bf (b)}.  The length and time units are
fixed by setting $R=1$ and $R^2/D = 1$. }
\label{fig:Tall}
\end{figure*}

Finally, Fig. \ref{fig:Tall} compares several mean reaction times for
the concentric domain.  One can see that the behaviour of the median,
the harmonic and the logarithmic means resembles that of the
conventional (arithmetic) mean FPT.  In particular, all these means
behave as $1/\kappa$ at small $\kappa$, indicating that the reaction
is limited by the kinetics.  Only the most probable FPT exhibits a
very different behaviour and shows almost no depedence on the
reactivity $\kappa$, as discussed above.

\subsection*{Reaction depth}

\begin{figure*}
\begin{center}
\includegraphics[width=18cm]{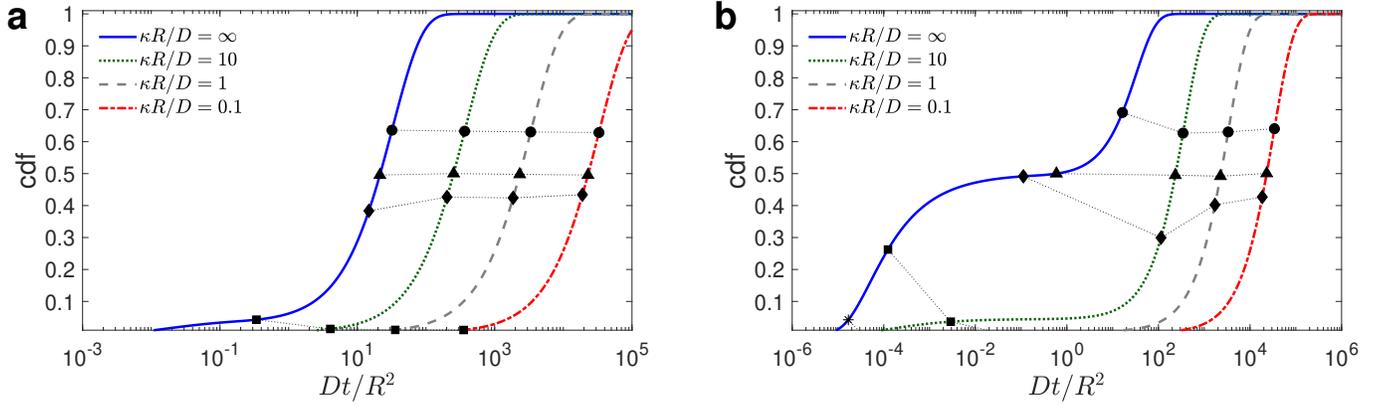} 
\end{center}
\caption{
Cumulative distribution function of reaction times, $F_r(t)$.  It is
evaluated for the reaction on an inner target of radius $\rho/R=0.01$,
with the starting point {\bf (a)} $r/R=0.2$ and {\bf (b)} $r/R=0.02$
and varying reactivity $\kappa$.  Symbols indicate the relevant
characteristic times: most probable time $t_{\rm mp}$ (asterisks),
harmonic mean $t_{\rm harm}$ (squares), logarithmic mean $t_{\rm log}$
(diamonds), median $t_{\rm median}$ (triangles), and mean time
(circles). Note that some most probable times are not seen at this
scale.}
\label{fig:ht}
\end{figure*}

The reaction depth \eqref{rdepth} is shown in figure \ref{fig:ht}.
Note first that the reaction depths corresponding to the shortest
characteristic time $t_{\rm mp}$ are evidently the shortest, amounting
to only about $4\%$ for perfect reactions and $r$ close to $\rho$.
For finite $\kappa$ or for starting points further away from the
target, the reaction depth $F_r(t_{\rm mp})$ diminishes.  In turn, in
all cases the reaction depth connected to the intermediate plateau is
dominant, increasingly so due to the reaction-control at lower
reactivities.

\section*{Data availability statement}

Data sharing not applicable to this article as no datasets were
generated or analysed during the current study.

\section*{Code availability statement} 

All figures have been prepared by means of Matlab software.  The
plotted quantities have been computed by explicit formulas provided in
the paper by using custom routines for Matlab software.  While the
explicit form makes these numerical computations straightforward,
custom routines are available from the corresponding author upon
request.

\acknowledgments

DSG acknowledges the support under Grant No. ANR-13-JSV5-0006-01 of the
French National Research Agency.  RM acknowledges funding from Deutsche
Forschungsgemeinschaft (project ME 1535/6-1) and from the Foundation for
Polish Science within a Humboldt Polish Honorary Research Scholarship.

\section*{Author contributions}

DSG, RM and GO formulated the problem.  DSG performed the mathematical
derivations and prepared the figures.  DSG, RM and GO analysed the
obtained results and wrote the manuscript.

\section*{Competing interests}

The authors declare no competing interests.

\section*{Materials \& Correspondence}

Correspondence and materials requests should be sent to D.~S.~Grebenkov,
denis.grebenkov@polytechnique.edu.


\begin{thebibliography}{40}

\bibitem{Calef83} 		Calef, D.F., and Deutch, J. M.
				Diffusion-controlled reactions, 
				{\it Ann. Rev. Phys. Chem.} {\bf 34}, 493--524 (1983).

\bibitem{Weiss86} 		Weiss, G. H. 
				Overview of theoretical models for reaction rates, 
				{\it J. Stat. Phys.} {\bf 42}, 3--36 (1986).

\bibitem{Shoup82} 		Shoup, D., and Szabo, A. 
				Role of diffusion in ligand binding to macromolecules and cell-bound receptors, 
				{\it Biophys. J.} {\bf 40}, 33--39 (1982).

\bibitem{Lindenberg}		Lindenberg, K., Oshanin, G., and Tachiya, M. (Eds)
				Reaction Kinetics Beyond the Textbook: Fluctuations, Many Particle Effects and Anomalous Dynamics,
				{\it J. Phys.: Condens. Matter} {\bf 19}, 1 (2007).

\bibitem{Hanggi90} 		H\"anggi, P., Talkner, P., and Borkovec, M. 
				Reaction-rate theory: fifty years after Kramers, 
				{\it Rev. Mod. Phys.} {\bf 62}, 251--341 (1990).

\bibitem{Krapivsky} 		Krapivsky, P. L., Redner, S., and Ben-Naim, E. 
				A Kinetic View of Statistical Physics 
				(Cambridge University Press, Cambridge, 2010).

\bibitem{Oshanin89a}		Oshanin, G., and Burlatsky, S. F. 
				Fluctuation-induced kinetics of reversible coagulation, 
				{\it J. Phys. A: Math. Gen.} {\bf 22}, L973--L976 (1989).

\bibitem{Oshanin89b}		Oshanin, G., Ovchinnikov, A. A., and Burlatsky, S. F. 
				Fluctuation-induced kinetics of reversible reactions,
				{\it J. Phys. A: Math. Gen.} {\bf 22}, L977--L982 (1989); 


\bibitem{Yuste08}		Yuste, S. B., Oshanin, G., Lindenberg, K., B\'enichou, O., and Klafter, J. 
				Survival probability of a particle in a sea of mobile traps: A tale of tails, 
				{\it Phys. Rev. E} {\bf 78}, 021105 (2008).

\bibitem{Grebenkov10a}  	Grebenkov, D. S., 
				Searching for partially reactive sites: Analytical results for spherical targets,
				{\it J. Chem. Phys.} {\bf 132}, 034104 (2010).

\bibitem{Grebenkov10b} 		Grebenkov, D. S.,
				Subdiffusion in a bounded domain with a partially absorbing-reflecting boundary, 
				{\it Phys. Rev. E} {\bf 81}, 021128 (2010). 


\bibitem{Smoluchowski1917}	Smoluchowski, M. 
				Versuch einer matematischen theorie der koagulationskinetik kolloider l\"osungen, 
				{\it Z. Phys. Chem.} {\bf 92}, 129--168 (1917).

\bibitem{Collins49} 		Collins, F. C., and Kimball, G. E.,  
				Diffusion-controlled reaction rates, 
				{\it J. Colloid Sci.} {\bf 4}, 425--437 (1949).

\bibitem{Sapoval94} 		Sapoval, B. 
				General Formulation of Laplacian Transfer Across Irregular Surfaces, 
				{\it Phys. Rev. Lett.} {\bf 73}, 3314--3316 (1994).

\bibitem{Grebenkov06} 		Grebenkov, D. S. 
				Partially Reflected Brownian Motion: A Stochastic Approach to Transport Phenomena, 
				in ``Focus on Probability Theory'', Ed. L. R. Velle, pp. 135--169 (Nova Science Publishers, New York, 2006). 


\bibitem{Holcman13} 		Holcman, D. and Schuss, Z.
				Control of flux by narrow passages and hidden targets in cellular biology, 
				{\it Phys. Progr. Rep.} {\bf 76}, 074601 (2013).

\bibitem{Gudowska17}		Gudowska-Nowak, E., Lindenberg, K., and Metzler, R. (Eds)
				Marian Smoluchowski's 1916 paper -- a century of inspiration,  
				{\it J. Phys. A: Math. Theor.} {\bf 50}, 380301 (2017).



\bibitem{Kolesov07} 		Kolesov, G., Wunderlich, Z., Laikova, O. N., Gelfand, M. S., and Mirny, L. A. 
				How gene order is influenced by the biophysics of transcription regulation, 
				{\it Proc. Natl. Acad. Sci. USA} \textbf{104}, 13948--13953 (2007).

\bibitem{Pulkkinen13} 		Pulkkinen, O., and Metzler, R. 
				Distance matters: the impact of gene proximity in bacterial gene regulation, 
				{\it Phys. Rev. Lett.} \textbf{110}, 198101 (2013).

\bibitem{Alberts} 		Alberts, B., Johnson, A., Lewis, J., Morgan, D., Raff, M., Roberts, K., and Walter, P., 
				{\it Molecular Biology of the Cell} 
				(Garland Science, New York, NY, 2014).

\bibitem{Snustad} 		Snustad, D. P., and Simmons, M. J. 
				{\it Principles of Genetics} 
				(Wiley, New York, 2000).

\bibitem{Godec16a} 		Godec, A., and Metzler, R.
				Universal Proximity Effect in Target Search Kinetics in the Few-Encounter Limit, 
				{\it Phys. Rev. X} {\bf 6}, 041037 (2016).

\bibitem{Godec16b} 		Godec, A., and Metzler, R.,
				First passage time distribution in heterogeneity controlled kinetics: going beyond the mean first passage time,
				{\it Sci. Rep.} {\bf 6}, 20349 (2016).


\bibitem{Mejia11} 		Mej\'{i}a-Monasterio, C., Oshanin, G., and Schehr, G. 
				First passages for a search by a swarm of independent random searchers, 
				{\it J. Stat. Mech.} P06022 (2011).

\bibitem{Mattos12} 		Mattos, T., Mej\'{i}a-Monasterio, C., Metzler, R., and Oshanin, G.
				First passages in bounded domains: When is the mean first passage time meaningful?, 
				{\it Phys. Rev. E} {\bf 86}, 031143 (2012).

\bibitem{Mattos14}		Mattos, T., Mej\'{i}a-Monasterio, C., Metzler, R., Oshanin, G., and Schehr, G. 
				Trajectory-to-trajectory fluctuations in first-passage phenomena in bounded domains, 
				in ``First-Passage Phenomena and Their Applications'', R. Metzler, G. Oshanin, and S. Redner (Eds)
				(Singapore, World Scientific, 2014).

\bibitem{Redner} 		Redner, S.,
				{\it A Guide to First Passage Processes}
				(Cambridge, Cambridge University press, 2001).


\bibitem{Benichou10} 		B\'enichou, O., Chevalier, C., Klafter, J., Meyer, B., and Voituriez, R.,
				Geometry-controlled kinetics, 
				{\it Nature Chem.} {\bf 2}, 472--477 (2010).

\bibitem{Benichou14} 		B\'enichou, O., and Voituriez, R.,
				From first-passage times of random walks in confinement to geometry-controlled kinetics, 
				{\it Phys. Rep.} {\bf 539}, 225--284 (2014).



\bibitem{Moffitt10} 		Moffitt, J. R., Chemla, Y. R., and Bustamante, C.  
				Mechanistic constraints from the substrate concentration dependence of enzymatic fluctuations, 
				{\it Proc. Natl. Acad. Sci. USA} {\bf 107}, 15739--15744 (2010).

\bibitem{Shoup81} 		Shoup, D., Lipari, G., and Szabo, A.
				Diffusion-controlled bimolecular reaction rates, 
				{\it Biophys. J.} \textbf{36}, 697--714 (1981).

\bibitem{Hippel89} 		von Hippel, P. H., and Berg, O. 
				Facilitated target location in biological systems, 
				{\it J. Biol. Chem.} \textbf{264}, 675--678 (1989).

\bibitem{Zon05} 		van Zon, J. S., and ten Wolde, P. R. 
				Simulating Biochemical Networks at the Particle Level and in Time and Space: Green's Function Reaction Dynamics, 
				{\it Phys. Rev. Lett.} \textbf{94}, 128103 (2005).



\bibitem{Grebenkov17a} 		Grebenkov, D. S., and Oshanin, G.
				Diffusive escape through a narrow opening: new insights into a classic problem, 
				{\it Phys. Chem. Chem. Phys.} {\bf 19}, 2723--2739 (2017).

\bibitem{Grebenkov17b} 		Grebenkov, D. S., Metzler, R., and Oshanin, G. 
				Effects of the target aspect ratio and intrinsic reactivity onto diffusive search in bounded domains, 
				{\it New J. Phys.} {\bf 19}, 103025 (2017).

\bibitem{Grebenkov2018} 	Grebenkov, D. S., Metzler, R. and Oshanin, G. 
				Towards a full quantitative description of single-molecule reaction kinetics in biological cells, 
				{\it Phys. Chem. Chem. Phys.} {\bf 20}, 16393--16401 (2018).




\bibitem{Gardiner} 		Gardiner, C. W., 
				{\it Handbook of stochastic methods for physics, chemistry and the natural sciences} 
				(Berlin, Springer, 1985).

\bibitem{Carslaw} 		Carslaw, H. S., and Jaeger, J. C.  
				{\it Conduction of Heat in Solids}
				(Oxford University Press, 1959).

\bibitem{Oshanin17} 		Oshanin, G., Popescu, M., and Dietrich, S.  
				Active colloids in the context of chemical kinetics,
				{\it J. Phys. A: Math. Theor.} {\bf 50}, 134001 (2017)

\bibitem{Evans11} 		Evans, M. R., and Majumdar, S. N. 
				Diffusion with stochastic resetting, 
				{\it Phys. Rev. Lett.} {\bf 106}, 160601 (2011).

\bibitem{Dean14} 		Dean, D. S., Gupta, S., Oshanin, G., Schehr, G., and Rosso, A.  
				Diffusion in periodic, correlated random forcing landscapes, 
				{\it J. Phys. A: Math. Theor.} {\bf 47}, 372001 (2014).

\bibitem{Meyer11} 		Meyer, B., Chevalier, C., Voituriez, R., and B\'enichou, O. 
				Universality classes of first-passage-time distribution in confined media,
				{\it Phys. Rev. E} {\bf 83}, 051116 (2011).

\bibitem{Isaacson13} 		Isaacson, S. A., and Newby, J. 
				Uniform asymptotic approximation of diffusion to a small target, 
				{\it Phys. Rev. E} {\bf 88}, 012820 (2013).


\end{thebibliography}
\end{document}